\documentclass{ws-mpla}
\usepackage[super]{cite}
\usepackage{amssymb}
\usepackage{amsfonts}
\usepackage{amsmath}
\usepackage{latexsym}
\usepackage[T1]{fontenc}
\usepackage[utf8]{inputenc}
\usepackage{lmodern}
\usepackage{graphicx}
\begin{document}

\markboth{K. Urbanowski}
{Remarks on the
 uncertainty relations}

\catchline{}{}{}{}{}

\title{REMARKS ON THE
 UNCERTAINTY RELATIONS
}

\author{\footnotesize KRZYSZTOF URBANOWSKI\footnote{K. Urbanowski}}

\address{University of Zielona G\'{o}ra, Institute of Physics, \\
ul. Prof. Z. Szafrana 4a, 65--516 Zielona G\'{o}ra, Poland \footnote{
University of Zielona G\'{o}ra, Institute of Physics,
ul. Prof. Z. Szafrana 4a, 65--516 Zielona G\'{o}ra, Poland; K.Urbanowski@if.uz.zgora.pl, $\;$ k.a.urbanowski@gmail.com}\\
K.Urbanowski@if.uz.zgora.pl, $\;$ k.a.urbanowski@gmail.com}

\maketitle

\pub{Received (Day Month Year)}{Revised (Day Month Year)}

\begin{abstract}
We analyze  general uncertainty relations and we show that
there can exist  such pairs  of non--commuting observables $A$ and $B$  and such
vectors that the lower bound for the product of standard deviations $\Delta A$ and  $\Delta B$ calculated for these vectors is zero: $\Delta A\,\cdot\,\Delta B \geq 0$.
We show also that for some pairs of non--commuting observables  the sets of vectors for which $\Delta A\,\cdot\,\Delta B \geq 0$ can be complete
(total). The Heisenberg, $\Delta  t \,\cdot\, \Delta E \geq \hbar/2$,
and Mandelstam--Tamm (MT), $ \tau_{A}\,\cdot \,\Delta E \geq \hbar/2$,  time--energy uncertainty relations
 ($\tau_{A}$ is the characteristic time for the observable $A$) are analyzed too. We show  that the interpretation  $\tau_{A} = \infty$ for eigenvectors of a Hamiltonian $H$ does not follow from the rigorous analysis of MT relation. We show also that contrary to the position--momentum uncertainty relation,
the validity of the MT relation is limited: It does not hold on complete sets of eigenvectors of $A$ and $H$.


\end{abstract}

\ccode{03.65.-w - Quantum mechanics; 03.65.Ta - Foundations of quantum mechanics; 01.55.+b - General physics}


\section{Introduction}
The famous Heisenberg uncertainty relations \cite{H,H2} play an important and significant  role in the understanding of the quantum world and in explanations of its properties.
There is a mathematically rigorous derivation of the position--momentum uncertainty relation
and the uncertainty relation for any pair of non--commuting observables, say $A$ and $B$,
but the same cannot be said about time--energy uncertainty relation. Nonetheless the time--energy uncertainty relation is considered by many authors using it as having the same status as the position--momentum uncertainty relation and it is often used  as the basis for drawing far--reaching conclusions regarding the prediction of the behavior
of some physical systems in certain situations in various areas of physics and astrophysics
and from time to time  such conclusions were considered as the crucial.
So, the time--uncertainty relation still  requires its analysis and checking whether it is correct and well motivated by postulates of quantum mechanics.
Here we present the analysis of the
general uncertainty relation derived by Robertson \cite{Robertson} and Schr\"{o}dinger \cite{Schrod-1930} and also of the
Heisenberg and Mandelstam--Tamm (MT) time--energy uncertainty  relations made within the framework of the standard formalism of Schrodinger and von Neumann quantum mechanics and we show that the validity of these time--energy uncertainty relations is limited.
In Section 2  the reader finds  a general theory and calculations. The MT time--energy uncertainty  relation is analyzed in Sec. 3.
    Discussion is presented in Sec. 4. Sec. 5 contains conclusions.

\section{Uncertainty principles }

The uncertainty principle belongs to one of the most characteristic and  important consequences of the quantum mechanics.
The most known form of this principle is the Heisenberg uncertainty principle \cite{H} for the position and momentum, which can be written as follows
\begin{equation}
\Delta_{\phi} x\ \cdot \Delta_{\phi} p_{x}\,\geq \,\frac{\hbar}{2}, \label{H1}
\end{equation}
where, according to Heisenberg, $\Delta_{\phi} x$ and $\Delta_{\phi} p_{x}$ are {\em "precisions"} with which the values $x$ and  $p$ are known \cite{H}.
The current interpretation of  $\Delta_{\phi} x$ and $\Delta_{\phi} p_{x}$ follows from the derivation of the uncertainty relation made by Robertson
\cite{Robertson} and Schr\"{o}dinger \cite{Schrod-1930},  (see also \cite{M}).  According  to them $\Delta_{\phi} x$ and $\Delta_{\phi} p_{x}$ denote the  standard (root--mean--square) deviations: In a general case for an observable $F$ the standard deviation is defined as follows
\begin{equation}
\Delta_{\phi} F = \| \delta F|\phi\rangle\|, \label{dF}
\end{equation}
where
$\delta F = (F - \langle F\rangle_{\phi}\,\mathbb{I} )$, and $\langle F\rangle_{\phi} \stackrel{\rm def}{=} \langle \phi|F|\phi\rangle$ is the average (or expected) value of an observable $F$ in a system whose state is represented by the normalized vector $|\phi\rangle \in {\cal H}$), provided that $|\langle\phi|F|\phi \rangle |< \infty$.
Equivalently:  $\Delta_{\phi} F \equiv \sqrt{\langle F^{2}\rangle_{\phi} - \langle F\rangle_{\phi}^{2}}$.
(In Eq. (\ref{H1})  $F$ stands for position  and momentum operators $x$ and $p_{x}$ as well as for their squares). The observable $F$ is represented by hermitian operator $F$ acting in a Hilbert space ${\cal H}$ of states $|\phi\rangle$. In general, the relation (\ref{H1}) results from basic assumptions of the quantum theory and from the geometry of Hilbert space \cite{Teschl}. Analogous relations hold for any two observables, say $A$ and $B$, represented by non--commuting hermitian operators $A$ and $B$ acting in the Hilbert space of states (see \cite{Robertson} and also \cite{Schrod-1930}), such that $[A,B]$ exists and $|\phi\rangle \in {\cal D}(AB) \bigcap {\cal D}(BA)$, (${\cal D}({\cal O})$ denotes the domain of an operator $\cal O$ or of a product of operators):
\begin{equation}
\Delta_{\phi} A \cdot \Delta_{\phi} B\;\geq\;\frac{1}{2} \left|\langle [A,B] \rangle_{\phi} \right|,\label{R1}
\end{equation}
where the equality takes place if $\left(B - \langle B\rangle_{\phi}\right)|\phi \rangle = i \lambda\left(A - \langle A\rangle_{\phi}\right)|\phi \rangle$, (here, $\lambda = \lambda^{\ast}$), or if $|\phi \rangle$ is an eigenvector  for operators $A$ or $B$, (see, eg. \cite{Teschl}). The derivation of  inequality (\ref{R1}) is the rigorous one.

Various derivations of the Heisenberg inequality can be  found in the literature and in  textbooks.
One of methods of deriving the uncertainty relation (\ref{R1}), which can be found in the literature, is the following:
The first step is to use the obvious relation resulting from the Schwartz inequality,
\begin{equation}
\left\| \delta A |\phi\rangle \right\|^{2}\;\left\| \delta B|\phi\rangle \right\|^{2}\,\geq \,\left|\langle\phi| \delta A\;\delta B|\phi \rangle \right|^{2}, \label{dAdB}
\end{equation}
which holds  for all $|\phi\rangle \in {\cal D}(AB) \bigcap {\cal D}(BA)$,
(where  $ (\Delta_{\phi} A)^{2} \equiv \left\|\delta A|\phi\rangle \right\|^{2}$ and $ (\Delta_{\phi} B)^{2} \equiv \left\|\delta B|\phi\rangle \right\|^{2}$). Next
step consists in a suitable transformation of the right hand side of Eq. (\ref{dAdB}),
\begin{eqnarray}
\left|\langle\phi| \delta A\;\delta B|\phi \rangle \right|^{2} & = & \left[ \Re\,(\langle \phi|\delta A\;\delta B|\phi \rangle) \right]^{2} +
\left[ \Im\,(\langle \phi|\delta A\;\delta B|\phi \rangle) \right]^{2}, \nonumber \\
&=& \frac{1}{4} \left(\langle\phi|( \delta A\;\delta B\,+\, \delta B\;\delta A)|\phi \rangle \right)^{2} \nonumber \\
&  &\;\;\;\; \;\;\;+\; \frac{1}{4} \left|\langle\phi|( \delta A\;\delta B\,-\, \delta B\;\delta A)|\phi \rangle \right|^{2} \nonumber \\
 &\equiv &
 \frac{1}{4} \left(\langle\phi|( \delta A\;\delta B\,+\, \delta B\;\delta A)|\phi \rangle \right)^{2} \,+\,
 \frac{1}{4} \left|\langle\phi|[A,B]|\phi \rangle \right|^{2}  \label{Sch-1}\\
 & \geq & \frac{1}{4} \left|\langle\phi|[A,B]|\phi \rangle \right|^{2}, \label{R1+1}
\end{eqnarray}
where $\Re\,(z)$ denotes the real part of the complex number $z$ and $\Im\,(z)$ is the imaginary part of $z$.
The property $[\delta A, \delta B] = [A,B]$  taking place  for all $|\phi\rangle \in {\cal D}(AB) \bigcap {\cal D}(BA)$ was used in (\ref{Sch-1}).

Now if one replaces the right hand side of Eq. (\ref{dAdB}) by (\ref{Sch-1}) then one obtains the uncertainty relation of the type derived by Schr\"{o}dinger \cite{Schrod-1930}:
\begin{equation}
 (\Delta_{\phi} A )^{2}\, \cdot\,( \Delta_{\phi} B)^{2}\,\geq
\frac{1}{4} \left(\langle\phi|( \delta A\;\delta B\,+\, \delta B\;\delta A)|\phi \rangle \right)^{2} \,+\,
 \frac{1}{4} \left|\langle\phi|[A,B]|\phi \rangle \right|^{2},  \label{Sch-2}
\end{equation}
or,   equivalently, in more familiar form,
\begin{equation}
 (\Delta_{\phi} A )^{2}\, \cdot\,( \Delta_{\phi} B)^{2}\,\geq
 \left(\frac{ \langle(AB +BA)\rangle_{\phi}}{2} -  \langle A\rangle_{\phi}\,\langle B \rangle_{\phi}\right)^{2} \,+\,
  \left|\frac{\langle [A,B] \rangle_{\phi}}{2} \right|^{2}.  \label{Sch-3}
\end{equation}
Note that relations (\ref{Sch-2}), (\ref{Sch-3})  seem to be  more general than the relation (\ref{R1}).
On the other hand, if  one replaces the right hand side in Eq. (\ref{dAdB}) by (\ref{R1+1}) then one obtains the uncertainty relation (\ref{R1})  as a result.

Let us analyze now the cases  of vanishing  expectation value of the commutator $[A,B]$, i.e., the cases  when $\langle\phi|[A,B]|\phi \rangle = 0$.
Note that it is not necessary for $A$ and $B$ to commute, $[A,B] =0$, in order that  $\langle\phi|[A,B]|\phi \rangle = 0$ for some $|\phi\rangle \in {\cal H}$. Simply it may happen that for some $|\phi \rangle \in {\cal H}$ and for some non-commuting observables $A$ and $B$ there is  $\langle\phi|[A,B]|\phi \rangle = 0$.

First, let us assume that there exist some non--commuting observables $A$ and $B$ and some vectors $|\phi\rangle \in {\cal H}$, which are not eigenvectors for  $A$ and $B$, such that $\langle\phi|[A,B]|\phi \rangle = 0$ for these $|\phi\rangle$. Then
\begin{equation}
\langle\phi|[A,B]|\phi \rangle = 0\;\;\Rightarrow\;\; \langle\phi|AB|\phi \rangle = \langle\phi |BA|\phi \rangle = (\langle\phi|AB|\phi \rangle)^{\ast}.
\label{[ab]=0}
\end{equation}
where generally $\langle \phi |AB|\phi \rangle \neq 0$.
The relation (\ref{[ab]=0}) implies that also $\langle\phi|\delta A \,\delta B|\phi \rangle$ $ = \langle\phi |\delta B \, \delta A|\phi \rangle = (\langle\phi|\delta A \,\delta B|\phi \rangle)^{\ast}$ and in such a case the uncertainty relations (\ref{Sch-2}), (\ref{Sch-3})  take the form of the output inequality  (\ref{dAdB}). The another situation is  in the case of the inequality (\ref{R1}):  If  it happens that  $\langle\phi|[A,B]|\phi \rangle = 0$ for  some $|\phi\rangle \in {\cal H}$ then for these $|\phi\rangle$ the inequality (\ref{R1}) takes the following form
\begin{equation}
\Delta_{\phi} A \cdot \Delta_{\phi} B\;\geq\;0, \label{dAdB=0}
\end{equation}
which means that in such a case, contrary to the inequalities (\ref{Sch-2}),  (\ref{Sch-3}),  the inequality (\ref{dAdB=0}) does not impose  any restrictions for the values of $\Delta_{\phi} A$ and $ \Delta_{\phi} B$.

Now let us consider the second case: The case
when $[A,B] \neq 0$ and $|\phi \rangle$ is normalized eigenvector of $A$ or $B$. So let $|\phi \rangle = |\phi_{\alpha}\rangle \in \Sigma_{A} \bigcap {\cal D}(B)$, where $\Sigma_{A} \subset {\cal H}$ is the set of all eigenvectors for $A$ (or  $|\phi \rangle = |\phi_{\beta}\rangle \in \Sigma_{B} \bigcap {\cal D}(A)$ and here   $\Sigma_{B} \subset {\cal H}$  denotes the set of all eigenvectors for $B$). We assume that sets $\Sigma_{A}, \Sigma_{B}$ are not empty. Typical observables are represented by self--adjoint operators, whose eigenvectors usually form a linearly dense (complete, or total) set in the Hilbert (state) space ${\cal H}$. Our analysis  refers to just such cases and does not apply to cases of observables with only continuous spectrum.
There is $A|\phi_{\alpha}\rangle = a_{\alpha}|\phi_{\alpha}\rangle$ for all $|\phi_{\alpha}\rangle \in \Sigma_{A}$ and therefore
$\langle\phi_{\alpha}|[A,B]|\phi_{\alpha} \rangle = 0$  and $\delta  A |\phi_{\alpha}\rangle = 0$ for all $|\phi_{\alpha}\rangle \in \Sigma_{A} \bigcap {\cal D}(A)$. This means that the right hand sides of the inequalities (\ref{R1}), (\ref{dAdB})  and (\ref{Sch-2}) take the zero values and all these inequalities take the form of the inequality (\ref{dAdB=0}).
(In fact  the property $\delta  A |\phi_{\alpha}\rangle = 0$ means that one obtains the result (\ref{dAdB=0}) directly from (\ref{dAdB})\,).
What is more  $\Delta_{\phi_{\alpha}}A = \|\delta A|\phi_{\alpha}\rangle\| = 0$  in the considered case, what together with  the relation (\ref{dAdB=0}) imply that in such a case the inequality (\ref{dAdB=0}) does not impose any restrictions on the standard (root–mean–square) deviation  $\Delta_{\phi_{\alpha}}B$ besides the condition that there should be  $0 \leq \Delta_{\phi_{\alpha}}B < \infty$.

One meets analogous situations when $|\phi\rangle = |\phi_{\beta}\rangle \in \Sigma_{B} \bigcap {\cal D}(A)$.
So, the considered case shows that there are situations (that is there are some state vectors $|\phi\rangle \in {\cal H}$) such that  the uncertainty relations (\ref{R1}), (\ref{Sch-2}), (\ref{Sch-3})  lead to no restrictions on  the standard deviations of non--commuting  observables $A$ and $B$. The set of vectors for which this property takes place needs not to be   small: Contrary, sets $\Sigma_{A}, \Sigma_{B}$,
as the sets of eigenvectors for self--adjoint operators,
are usually linearly dense (complete) sets in the state space ${\cal H}$ and can be used as the basis in ${\cal H}$.

\section{Time--energy uncertainty relations }

Now we can apply  results of the previous Section to the analyze of the time--energy uncertainty relations.
The validity of the analogous relation to (1) for the time and energy was postulated by Heisenberg in \cite{H} (see also \cite{J}).
This time--energy uncertainty relation was a result of Heisenberg's heuristic considerations
and it is usually written as follows
\begin{equation}
\Delta_{\phi} t \cdot \Delta_{\phi} E \geq \frac{ \hbar}{2}.\label{H2}
\end{equation}
According to Heisenberg, this relation {\em "shows how a precise determination of the energy can only be obtained at the cost of a corresponding uncertainty in the time"} \cite{H}.

The more rigorous derivation
of the inequality of the form (\ref{H2}) was proposed
 by Mandelstm and Tamm \cite{M-T} and now it is known as the Mandelstam--Tamm time--energy uncertainty relation. Their derivation is reproduced in \cite{M}.
 The starting point of this derivation is the general uncertainty relation  (\ref{R1}).
In  (\ref{R1}) the operator $B$ is replaced by the selfadjoint non--depending on time Hamiltonian $H$ of the system considered and $\Delta_{\phi} B$ is replaced by $\Delta_{\phi} H $ and then identifying the standard deviation $\Delta_{\phi} H $ with $\Delta_{\phi} E$ one finds that
\begin{equation}
\Delta_{\phi} A \cdot \Delta_{\phi} E\;\geq\;\frac{1}{2} \left| \langle [A,H] \rangle_{\phi} \right|,\label{M1}
\end{equation}
where it is assumed that $A$ does not depend upon the time $t$ explicitly, $|\phi\rangle \in {\cal D}(HA) \bigcap {\cal D}(AH)$,  and $[A,H]$ exists.
The next step is to use the Heisenberg representation and corresponding equation of motion which allows
to replace the average value of the commutator standing in the right--hand side of the inequality (\ref{M1}) by the derivative with respect to  time $t$ of the expected value of $A$,
\begin{equation}
\langle [A,H] \rangle_{\phi}  \equiv i\hbar  \frac{d}{dt} \langle A \rangle_{\phi}. \label{M2}
\end{equation}
The Eq. (\ref{M2}) means the the inequality  (\ref{M1})
takes the following equivalent form,
\begin{equation}
\Delta_{\phi} A \cdot \Delta_{\phi} E\;\geq\;\frac{\hbar }{2} \left|  \frac{d}{dt} \langle A \rangle_{\phi} \right|.\label{M3}
\end{equation}
(Relations (\ref{M1}) --- (\ref{M3}) are rigorous). Next authors \cite{M,M-T} and many others divide both sides of the inequality (\ref{M3}) by the term $\left|  \frac{d}{dt} \langle A \rangle_{\phi} \right|$, which leads to the following relation
\begin{equation}
\frac{\Delta_{\phi} A \cdot  \Delta_{\phi} E }{\left|  \frac{d}{dt} \langle A \rangle_{\phi} \right|} \;  \geq \; \frac{\hbar}{2}, \label{M4}
\end{equation}
usually written as
\begin{equation}
\frac{\Delta_{\phi} A }{\left|  \frac{d}{dt} \langle A \rangle_{\phi} \right|}  \cdot  \Delta_{\phi} E\;  \geq \; \frac{\hbar}{2}, \label{M4a}
\end{equation}
or, using
\begin{equation}
 \tau_{A} \stackrel{\rm def}{=} \frac{\Delta_{\phi} A}{\left|  \frac{d}{dt} \langle A \rangle_{\phi} \right|},\label{tau}
 \end{equation}
 to the final result known as the MTtime--energy uncertainty relation,
 \begin{equation}
 \tau_{A} \cdot \Delta_{\phi} E \geq \frac{\hbar}{2}, \label{M5}
 \end{equation}
 where $\tau_{A}$ is usually considered as a time characteristic of the evolution of the statistic distribution of $A$ \cite{M}. The time--energy uncertainty relation (\ref{M5}) and the above described derivation of this relation is accepted by many authors analyzing this problem or applying this relation (see, e.g. \cite{JB,Bauer,Gislason,skr} and many other papers). On the other hand there are some formal controversies
regarding the role and importance of the  parameter $\tau_{A}$ in (\ref{M5}) or $\Delta t$ in (\ref{H2}). These controversies are caused by the fact that in the quantum mechanics the time $t$ is a parameter. Simply it  cannot be described by the self--adjoint operator, say $T$,  acting in the Hilbert space of states (that is time cannot be an observable) such that $[T,H] = i\hbar \mathbb{I}$ if the Hamiltonian $H$ is bounded from below.
This observation was formulated by Pauli \cite{Pauli} and it is know as "Pauli's Theorem" (see, eg. \cite{JB,Busch}). Therefore  the status of the relations (\ref{H2})
and relations (\ref{H1}), (\ref{R1}) is not the same
regarding the basic principles of the quantum theory (see also discussion, e.g.,  in \cite{Vor,Hi1,Hi2,Br}).

At this point it should be mentioned there was many attempts to derive the time--energy uncertainty relation using the "time" operator $T$. For example,
the relation $[T,H] = i\hbar \mathbb{I}$, (where the operator $T$ is self--adjoint), was used to derive rigorously the time--energy uncertainty relation from a quantum theory of events but one should remember that the {\em "events" theory} is not a "particle" theory like standard quantum theory and $H$ is not a "normal" quantum--mechanical Hamiltonian (see \cite{Edwards}). An another attempt to derive the inequality  (\ref{H2}) is related to a use of so--called
"{\em tempus}" operator $\cal T$ (see \cite{Kobe2,Kobe3}). Simply using in classical mechanics canonical transformation with suitable generating function $S(q,E,t)$ one can transform the "old" canonical variables $(q,p)$, (where $q$ is the position and $p$ is the momentum), to  a set of new variables $(q\,',p\,')$, such that the new canonical momentum equals to the energy $p\,'=E$ and the new generalized coordinate $q\,'$ conjugate to $p\,'=E$  has a dimension of the time  and it is denoted as  $\cal T$ and called {\em tempus}, $q\,'= {\cal T}$, (see \cite{Kobe1}). In general, tempus ${\cal T}= {\cal T}(q,E,t)$ is
a function of the old generalized coordinate $q$, the new canonical
momentum $p\,'=E$ and the time of evolution $t$. It is not unique because the
arbitrary function of the energy can be added to tempus \cite{Kobe2,Kobe3}. Next step performed in \cite{Kobe2,Kobe3} is to use the property that the energy and the tempus are canonically conjugate variables and must satisfy the same Poisson bracket as the "old" canonical variables $q$ and $p$: $\{E,{\cal T} \}\equiv \{q,p\} = 1$. This property is the basis in \cite{Kobe2,Kobe3} for the conclusion that replacing the tempus $\cal T$ and the energy $E$ by  hermitian operators $\hat{{\cal T}}$ and $\hat{ E}$ one can replace the Poisson bracket $\{E,{\cal T} \}$ by a commutator $[{\hat{E}},\hat{{\cal T}}]$ to obtain  for $\hat{E}, \hat{\cal T}$ the same commutation relation $[\hat{E},\hat{ {\cal T} }]= i \hbar \mathbb{I}$ as for the pair of position and momentum operators $q$ and $p$:\;$[q,p]= i\hbar \mathbb{I}$.
The commutation relation $[\hat{E},\hat{ {\cal T} }]= i \hbar \mathbb{I}$ obtained in this way
is used in \cite{Kobe2} to derive the inequality $\Delta E \cdot \Delta {\cal T} \geq \frac{\hbar}{2}$. At first glance, everything looks good but a more detailed analysis shows a lot of inconsistencies in the  approach used to derive this "uncertainty" relation. The main problems are the following:
The first unclear problem is that $\cal T$ is not unique, therefore the tempus operator $\hat{{\cal T}}$ is also not unique and in a result  the standard deviation $\delta {\cal T}$ needs not be unique. The second obscureness is connected with the dependence of the tempus $\cal T$  (and thus the operator $\hat{{\cal T}}$ too) on $q,E$ and the time $t$. Therefore $\Delta {\cal T}$ should also depend on these parameters:  $\Delta {\cal T} =  \Delta {\cal T}(q,E,t)$, which makes the interpretation of the inequality  $\Delta E \cdot \Delta {\cal T} \geq \frac{\hbar}{2}$ unclear. The biggest problem is the obscureness associated with the postulated commutation relation $[\hat{E},\hat{ {\cal T} }]= i \hbar \mathbb{I}$: There is no proof in  papers cited in \cite{Kobe2,Kobe3} and in references those one can find therein that the quantum theory resulting form  this commutation relation is  a unitary equivalent to the quantum theory resulting from the postulate that $[q,p]= i\hbar \mathbb{I}$. In order to prove that these quantum theories are equivalent one should prove that there exists such a unitary operator, say $R$, that $RpR^{-1} = p\,'\equiv \hat{E}$ and $RqR^{-1} = q\,'\equiv \hat{{\cal T}}(q,E,t)$. (If fact it is sufficient that $R$ is invertible). Without the mentioned proof the interpretation of the relation $\Delta E \cdot \Delta {\cal T} \geq \frac{\hbar}{2}$ is unclear.
In addition to the above reservations, one should remember the conclusions resulting from Paulie's theorem: if $\hat{E}$ is bounded from below, then $\hat{{\cal T}}$ cannot be self--adjoint.

The MT uncertainty relation (\ref{M5}) is also not free of controversies.
Researchers applying and using
the above described derivation of (\ref{M5})
in their discussions of the time-energy uncertainty relation
made use  (consciously or not) of a presumption that the
right hand sides of Es. (\ref{M1}), (\ref{M3}) are non--zero, that is that there does not exist any vector $|\phi\rangle \in {\cal H}$ such that $\langle[A,H]\rangle_{\phi}  = 0$,  or $d/dt\langle A\rangle_{\phi} =0$. Although in the original paper of Mandelstam and Tamm \cite{M-T} there is a reservation that for the validity of the formula of the type (\ref{M5}) it is necessary that $\Delta_{\phi} H \neq 0$ (see also, e.g. \cite{Gray,Aharonov}),
there are no analogous reservations in \cite{M} and in many other papers.

Basic principles  of mathematics require
that before  dividing  both sides of Eq. (\ref{M3})  by $\left|  \frac{d}{dt} \langle A \rangle_{\phi} \right|$, one should check whether $ \frac{d}{dt} \langle A \rangle_{\phi} $ is different from zero or not. Let us do this now: Let $ \Sigma_{H} \subset {\cal H}$  be a set of eigenvectors $ |\phi_{\beta}\rangle $ of $H$ for the eigenvalues $E_{\beta}$. Then, as it has been shown in the general case in the previous Section, there is
$H|\phi_{\beta}\rangle = E_{\beta}|\phi_{\beta}\rangle$
for all $|\phi_{\beta}\rangle \in \Sigma_{H}$ and therefore for all $|\phi_{\beta}\rangle \in \Sigma_{H}  \bigcap {\cal D}(A)$ (see (\ref{M2})),
\begin{equation}
\langle [A,H] \rangle_{\phi_{\beta}}  = i\hbar  \frac{d}{dt} \langle A \rangle_{\phi_{\beta}} \equiv 0. \label{U1}
\end{equation}
Similarly,
\begin{equation}
\Delta_{\phi_{\beta}} H = \sqrt{\langle| H^{2}|\rangle_{\phi_{\beta}} - (\langle |H|\rangle_{\phi_{\beta}})^{2}} \stackrel{\rm def}{=} \Delta_{\phi_{\beta}} E \equiv 0,
\label{U2}
\end{equation}
for all $|\phi_{\beta}\rangle \in \Sigma_{H}$. This means that in all such cases the non--strict inequality (\ref{M3}) takes
the form of the following equality
\begin{equation}
\Delta_{\phi} A \cdot 0 \;= \;\frac{\hbar }{2} \cdot 0, \label{U3}
\end{equation}
which is the particular case of the general result (\ref{dAdB=0}) obtained in the previous Section.
In other words, one cannot divide the both sides of the inequality (\ref{M3}) by $ \left|\frac{d}{dt} \langle A \rangle_{\phi_{\beta}}\right| \equiv 0 $  for all $|\phi_{\beta}\rangle \in \Sigma_{H}$, because in all such cases the result is  an undefined number and such mathematical operations are unacceptable.
It should be noted that although the authors of the publications \cite{M,Gray} knew that the property (\ref{U1}) occurs for the vectors from the set $\Sigma_{H}$, it did not prevent them to divide both sides of inequality equality (\ref{M3})  by  $\left|  \frac{d}{dt} \langle A \rangle_{\phi} \right|$, that is by  $\left|  \frac{d}{dt} \langle A \rangle_{\phi} \right| \equiv 0$,  without taking into account (\ref{U2}) and without any explanations.
What is more, this shows that there is no reason to think of $\tau_{A}$ as infinity in this case as it was done, e.g, in \cite{M,Gray}.
For example, in \cite{M} one can read at the end of \S  13, Chap. VIII:
{\em "If, in particular, the system is in a stationary state, $\frac{d}{dt} \langle A \rangle_{\phi} = 0$,
no matter what $ A$, and consequently $\tau_{A}$ is infinite; however, $\Delta E_{\phi} = 0$,
in conformity with relation (\ref{M5})"}. Note that this is exactly the case described by (\ref{U3}). Similar point of view one can meet, eg., in \cite{Gray}.  Our definition (\ref{tau}) of $\tau_{A}$ corresponds with the formula (11) in \cite{Gray} (our $\phi$ is replaced by $\psi$ in \cite{Gray}) and directly after this formula one reads: {\em "If $\psi$  is an eigenvector of $H$, then the denominator in (11) is always zero, thus no observable varies
in time. Thinking of $\tau_{A}$ as infinity in this case makes sense."} Again it is exactly the case described by our relation (\ref{U3}). Analogous statements to those cited one can meet in many other papers and books.

Summing up, the interpretation of the case $ \frac{d}{dt} \langle A \rangle_{\phi_{\beta}} \equiv 0$ as $\tau_{A} = \infty$ for $|\phi_{\beta}\rangle  \in \Sigma_{H}$ (or $|\phi_{\alpha}\rangle \in \Sigma_{A}$) is not a mathematical consequence of the relation (\ref{M3})
and of the derivation of the inequalities (\ref{M4}) and (\ref{M4a}). It is because if $ i\hbar \frac{d}{dt} \langle A \rangle_{\phi_{\beta}} \equiv \langle[A,H]\rangle_{\phi_{\beta}} = 0$ then always $\Delta_{\phi} H =  \Delta_{\phi} E =0$, which means that in this case the left hand side of (\ref{M4}) and  (\ref{M4a})
become an indefinite number. So, the statement that $\tau_{A} = \infty$ when $ \frac{d}{dt} \langle A \rangle_{\phi_{\beta}} = 0$ seems to be rather heuristic statement arbitrarily entered by hand. It should be noted that
none of the authors using this interpretation of $\tau_{A}$ evaluated the number of vectors for which condition (\ref{U1}) occurs and the size of the set of such vectors.
In general, the problem is that usually the set $\Sigma_{H}$ of the eigenvectors of the Hamiltonian $H$ is a linearly dense (complete) set in the
 state space ${\cal H}$. Hence the conclusion, that such relations as (\ref{M4a}) and then (\ref{M5}) are correct only for some
specific states $|\phi\rangle$ and observables $A$, and for others need not to be correct, seems to be valid and justified.

The following analysis confirms this conclusion. As it has been shown $\tau_{A}$ cannot be defined correctly for eigenvectors $|\phi_{\beta}\rangle $ of $H$, although $[A, H] \neq 0$. Let us consider now vectors close to eigenvectors $|\phi_{\beta}\rangle$. Defining
\begin{equation}
|\psi_{\eta} \rangle = N_{\eta}\left(|\phi_{\beta}\rangle + \eta |\psi\rangle \right), \label{psi-eta}
\end{equation}
(where $\eta$ is the real number, $N_{\eta}$ is the normalization constant: $\langle \psi_{\eta}|\psi_{\eta}\rangle = 1$,  $|\psi\rangle \neq |\phi_{\beta}\rangle$, $\langle \phi_{\beta}|\phi_{\beta}\rangle = \langle \psi|\psi\rangle = 1$, and $|\psi\rangle$ is not an eigenvector for $H$ and for $A$),
one can see that the distance $ d(\phi_{\beta},\psi_{\eta}) =\| |\phi_{\beta}\rangle  - |\psi_{\eta}\rangle \| $  tends to
zero when $\eta \to 0$. It is because $N_{\eta} \to 1$ when $\eta \to 0$. So, for $\eta \to 0$
vector $|\psi_{\eta}\rangle$ gets closer and closer to the vector $|\phi_{\beta}\rangle$.

There are  $H|\phi_{\beta}\rangle = E_{\beta}|\phi_{\beta}\rangle$, $\Delta_{\psi_{\eta}}A  = \|\delta A |\psi_{\eta}\rangle \| \neq 0$, $\Delta_{\psi_{\eta}}E \equiv \Delta_{\psi_{\eta}} H  = \|\delta H| \psi_{\eta}\rangle \| \neq 0$ and $\langle [A,H] \rangle_{\psi_{\eta}}  = i\hbar  \frac{d}{dt} \langle A \rangle_{\psi_{\eta}} \neq 0$ for $\eta \neq 0$, where $\delta A |\psi_{\eta}\rangle = (A - \langle A \rangle_{\psi_{\eta}})|\psi_{\eta}\rangle$ ($\delta H|\psi_{\eta}\rangle $ is calculated analogously). This means that the left hand side of the inequalities (\ref{M4}), (\ref{M4a}) are well defined and thus
$\tau_{A}^{\eta}$ calculated for $|\psi_{\eta}\rangle$ using the definition (\ref{tau}) is well defined too and finite which means that in this case the relation (\ref{M5}) is correct.
Then we can observe that $\lim_{\eta \to 0} \Delta_{\psi_{\eta}}A \neq 0$, but $\lim_{\eta \to 0} \Delta_{\psi_{\eta}}H = 0$ and $\lim_{\eta \to 0} \langle [A,H] \rangle_{\psi_{\eta}}  = i\hbar \, \lim_{\eta \to 0} \frac{d}{dt} \langle A \rangle_{\psi_{\eta}} = 0$,
which shows that in limiting case $\eta \to 0$ reservations leading to the Eq. (\ref{U3}) seem to be not removed. However a  more detailed analysis of  the Eqs (\ref{M4a}), (\ref{tau})  shows that
\begin{equation}
\tau_{A} = \lim_{\eta \to 0} \tau_{A}^{\eta} = \lim_{\eta \to 0}\, \frac{\Delta_{\psi_{\eta}} A }{\left|  \frac{d}{dt} \langle A \rangle_{\psi_{\eta}} \right|} =
\lim_{\eta \to 0}\, \hbar \frac{\Delta_{\psi_{\eta}} A }{\left| \langle [A,H] \rangle_{\psi_{\beta}} \right|} = \infty.
\label{eta1}
\end{equation}
Let us consider now the case of $|\phi\rangle = |\phi_{\alpha}\rangle$ being an  eigenvector for $A$.
(This case was also noticed in \cite{Gray}). Then also for any $|\phi_{\alpha}\rangle \in \Sigma_{A}\bigcap{\cal D}(H)$, (where by $\Sigma_{A}$ we denote the set of eigenvectors $|\phi_{\alpha}\rangle$ for $A$), $\left|  \frac{d}{dt} \langle A \rangle_{\phi} \right| \equiv 0$ and $\Delta_{\phi}A \equiv 0$. Thus, instead of (\ref{U3}) one  once more has
$\;\;0\cdot \Delta_{\phi}H = \frac{\hbar}{2}\cdot 0$, and once again dividing both sides of this inequality  by  zero has no mathematical sense.
Now  note that the relations (\ref{H1}), (\ref{R1}) are always  satisfied  for   all $|\phi\rangle \in {\cal H}$ fulfilling the conditions specified before Eq. (\ref{R1}),
and in contrast to this property, we have  proved that the MT relation (\ref{M4a}) may not be true not only on the set $\Sigma_{H} \subset {\cal H}$, whose  span is usually dense in ${\cal H}$,  but also on the set   $\Sigma_{A} \subset {\cal H}$.
So the conclusion  that for  eigenvectors $|\phi_{\alpha}\rangle \in \Sigma_{A} \subset {\cal H}$ the uncertainty relation (\ref{M4a}) may not be true seems to be justified.
In this context the following question arises: What can the result be if to consider the vectors close to these eigenvectors? The possible answer for this question can be found
performing similar analysis to that leading to  the result (\ref{eta1}). In order to this one can use similar vectors to those defined in Eq. (\ref{psi-eta}):
\begin{equation}
|\psi_{\lambda} \rangle = N_{\lambda}\left(|\phi_{\alpha}\rangle + \lambda |\psi\rangle \right), \label{psi-ka}
\end{equation}
(where $\lambda$ is the real number, $N_{\lambda}$ is the normalization constant,  $|\psi\rangle \neq |\phi_{\alpha}\rangle$, and $|\psi\rangle$ is not an eigenvector for $A$ and for $H$, $A |\phi_{\alpha}\rangle =  a_{\alpha}|\phi_{\alpha}\rangle$).
The distance $ d(\phi_{\alpha},\psi_{\lambda}) =\| |\phi_{\alpha}\rangle  - |\psi_{\lambda}\rangle \| $  tends to
zero when $\lambda \to 0$.  This means that for $\lambda \to 0$
vector $|\psi_{\lambda}\rangle$ gets closer and closer to the vector $|\phi_{\alpha}\rangle$. Here the situation is similar to the one leading to the result (\ref{eta1}):
We have $\Delta_{\psi_{\lambda}}A  = \|\delta A |\psi_{\lambda}\rangle \| \neq 0$, $\Delta_{\psi_{\lambda}}H   = \|\delta H| \psi_{\lambda}\rangle \| \neq  0$ and $\langle [A,H] \rangle_{\psi_{\lambda}}  = i\hbar  \frac{d}{dt} \langle A \rangle_{\psi_{\lambda}} \neq 0$ for $\lambda \neq 0$, where $\delta A |\psi_{\lambda}\rangle = (A - \langle A \rangle_{\psi_{\lambda}})|\psi_{\lambda}\rangle$ and so on. So also in this case
$\tau_{A} \equiv \tau_{A}^{\lambda}$ calculated for $|\psi_{\lambda}\rangle$ using the definition (\ref{tau}) is well defined and finite and therefore   the relation (\ref{M5}) is correct.
In the limiting case $ \lambda \to 0$ we have: $\lim_{\lambda \to 0} \Delta_{\psi_{\lambda}}A = 0$,  $\lim_{\lambda \to 0} \Delta_{\psi_{\lambda}}H \neq  0$ and $\lim_{\lambda \to 0} \langle [A,H] \rangle_{\psi_{\lambda}}  = i\hbar  \lim_{\lambda \to 0} \frac{d}{dt} \langle A \rangle_{\psi_{\lambda}} = 0$, which shows that in this case
doubts concerning the inequality (\ref{M5}) still hold. What is more using $\tau_{A}^{\lambda}$,
\begin{equation}
 \tau_{A}^{\lambda} \stackrel{\rm def}{=} \frac{\Delta_{\psi_{\lambda}} A} {\left|  \frac{d}{dt} \langle A \rangle_{\psi_{\lambda}} \right| } \equiv
 \hbar \frac{\Delta_{\psi_{\lambda}} A }{\left| \langle [A,H] \rangle_{\psi_{\beta}} \right|},
\label{ka1}
\end{equation}
one finds that contrary to the case of eigenvectors $|\phi_{\beta}\rangle$ for $H$ considered earlier the limit $\tau_{A} = \lim_{\lambda \to 0} \tau_{A}^{\lambda}$ is not unique and it depends on the choice of $|\psi\rangle$ in (\ref{psi-ka}).
This means that in the  case of  eigenvectors $|\phi_{\alpha}\rangle$ for $A$
all doubts concerning the definition and interpretation of $\tau_{A} $ remains and  that the MT relation (\ref{M4a}) does not apply in this case.

Let us come back for a moment to the case of eigenvectors for $H$ and the result (\ref{eta1}).
Mathematical correctness requires that looking for the limit $\eta \to 0$ of the left hand side of the inequality (\ref{M4}) calculated for $|\psi_{\eta}\rangle$ one can not consider only the fraction used in Eq (\ref{eta1}) but one
should calculate the limit $\eta \to 0$ of the full fraction $\left(\|\delta A|\psi_{\eta}\rangle\|\cdot\|\delta H|\psi_{\eta}\rangle\| \right)/\left| \frac{d}{dt} \langle A \rangle_{\psi_{\eta}} \right|  $. It is so because the inequality (\ref{M4}) is the result of dividing of two sides of the inequality (\ref{M3}) by $\left|\langle\frac{d}{dt} \langle A \rangle_{\psi_{\eta}} \right|$ and, contrary to relation (\ref{M4a}) is mathematically correct. (There are no mathematical reasons  to write (\ref{M4}) in the form (\ref{M4a}): It is an  arbitrary choice of many authors studying this problem).
The result of this limit
is similar to that obtained for the limit of $\tau_{A}^{\lambda}$ and it is non--unique:
\begin{equation}
\lim_{\eta \to 0}\, \frac{\|\delta A|\psi_{\eta}\rangle\|^{2}\;\|\delta H|\psi_{\eta}\rangle\|^{2}}{\left|  \frac{d}{dt} \langle A \rangle_{\psi_{\eta}} \right|^{2}} =
\lim_{\eta \to 0}\, \hbar^{2} \frac{\|\delta A|\psi_{\eta}\rangle\|^{2}\;\|\delta H|\psi_{\eta}\rangle\|^{2} }{\left| \langle [A,H] \rangle_{\psi_{\eta}} \right|^{2}} = c_{\psi}^{2},  \label{c-psi}
\end{equation}
where the value of $c_{\psi}$ depends on the choice of $|\psi\rangle$ in (\ref{psi-eta}). Taking this into account one should be careful when interpreting the result (\ref{eta1}).
Namely, the result (\ref{eta1}) suggests that  if to consider $\tau_{A}$ defined by Eq. (\ref{tau}) as the separate, independent quantity, then
in the limiting case  $|\psi_{\eta} \rangle  \underset{\eta \to 0}{\rightarrow}  |\phi_{\beta}\rangle$
the fraction (\ref{tau}) defining $\tau_{A}$ leads to the  acceptable result  $\tau_{A} = \lim_{\eta \to 0} \tau_{A}^{\eta}= \infty$ and thus the  inequality (\ref{M5}) with $\tau_{A}$ given by Eq. (\ref{eta1}) and $\lim_{\eta \to 0} \Delta_{\psi_{\eta}}H = 0 \equiv \Delta E_{\phi_{\beta}} = 0$
gives an impression to be correct.
On the other hand there is no unique limit of the left hand side of the inequality (\ref{M4}), when it is calculated for vectors of the type (\ref{psi-eta}).
So, although the result (\ref{eta1}) seems to be acceptable, the result (\ref{c-psi}) suggests that the assumption of its correctness may be wrong.
From this analysis one can conclude that even if for eigenvectors $|\phi_{\beta}\rangle $ of $H$ doubts concerning the inequality (\ref{M5}) and the interpretation of $\tau_{A}$ remain then for vectors infinitely close to eigenvectors $|\phi_{\beta}\rangle$ of $H$ everything is correct.
The problem is that in the literature and in applications
no one uses the inequality (\ref{M5}) with $\tau_{A}$ calculated for vectors close (or infinitely close) to eigenvectors $|\phi_{\beta}\rangle$ of $H$ and no one considers the inequality (\ref{M5}) and the case $\tau_{A} = \infty$
as the limiting case of the type described above.
Instead of this, $\tau_{A}$ is considered as the quantity defined separately and independently of Eqs (\ref{M3}), (\ref{M4}). In such a situation
the natural and acceptable conclusion is that for eigenvectors $|\phi_{\beta}\rangle$ of $H$ one has $\tau_{A} = \infty$.
Unfortunately from the mathematical point of view such thinking is wrong. The definition (\ref{tau}) of $\tau_{A}$ is strictly connected with relations (\ref{M3}), (\ref{M4}): The inequality (\ref{M4}) is the result of dividing of both sides of (\ref{M3}) by  $\left|\langle\frac{d}{dt} \langle A \rangle_{\psi_{\eta}} \right|$ which equals zero for eigenvectors of $A$ or $H$.
In a general case,
when the relation (\ref{M5}) is used one usually calculates standard deviations $\Delta_{\psi}A, \Delta_{\psi} H = \Delta_{\psi}E$ and $\tau_{A}$ for a given vector $|\psi\rangle$ but does not for vectors close to it. The same was done in \cite{M,M-T}. In fact problems with the MT uncertainty relation (\ref{M5}) for eigenvectors $|\phi_{\alpha}\rangle \in \Sigma_{A}$ and $|\phi_{\beta}\rangle \in \Sigma_{H}$ are  consequences of the result (\ref{dAdB=0}):
In such a case the inequalities (\ref{M1}), (\ref{M3}) take a form of (\ref{dAdB=0}), where there is no room for defining such quantities as $\tau_{A}$.
So, the intuitive thinking of $\tau_{A}$ as infinity in the case of stationary states is not based on conclusions resulting from the rigorous derivation of the relations (\ref{M4}), (\ref{M5}). Such an interpretation of $\tau_{A}$ makes a sense only
when one considers $\tau_{A}$ as the independently defined quantity being
 a limit of sequences of vectors approaching stationary states.

At this point it should be emphasized that
in this paper we discuss results obtained within
the approach based on the ideas presented in \cite{Robertson,Schrod-1930,M,Teschl} and others where the standard deviations are calculated for  pure quantum states, that is for given state vectors, but not for sets or sequences of vectors and therefore the interpretation of $\tau_{A}$ in (\ref{M5}) as the limiting case of the type
(\ref{eta1}) does not fall within this approach.

\section{Discussion}

As it was mentioned, uncertainty relations  calculated for a pure state corresponding to an eigenvalue from the discrete spectra of considered non--commuting observables
were analyzed in previous Sections.
(Mixed states were not considered). With such  assumptions there is no room to consider  sets (or sequences) of vectors tending to the given state.
It seems that the use of mixed states and the density matrix $\rho$ describing a state of the system
together with appropriate  uncertainty relation written for the  density matrix (see, e.g \cite{M-T,Dodonov}) can help in  finding a solution to
the problem of the limits of sequences of states tending to a given eigenvector,
or at least to work around this problem.

Referring to the density matrix and the uncertainty relation written using the density matrix
it should be noted however that the problem of dividing "zero" by "zero" was ignored in many papers, where the time--energy uncertainty relation for mixed states was considered.
One of the typical situation of this type one can meet, e.g. in \cite{Beretta}: For example: if to analyze formula (41) in  \cite{Beretta} we can  see that  assuming that $\rho$ is formed from the eigenvectors of the considered operator $F$, we  encounter the problem of dividing zero by zero described in Sec. 3. It is easy to see if to choose $\rho =
|\phi_{F}\rangle \langle \phi_{F}|$, where $|\phi_{F}\rangle$ is a normalized eigenvector for $F$.
Nevertheless, acting analogously as it is described in the previous Section, we can use the matrix
$\rho$ built from the normalized vectors of type (\ref{psi-ka}): $|\psi_{F,\,\epsilon}\rangle = N_{\epsilon}(|\phi_{F}\rangle + \epsilon |\psi\rangle)$ and then we can see that the operator defined by Eq. (41) has a the unit norm for $\epsilon \neq 0$ and
in the limit $\epsilon \to 0$ this norm will be also a unit norm in topology used in \cite{Beretta}.

Note also that the use of the Schr\"{o}dinger uncertainty relation (\ref{Sch-2}), (\ref{Sch-3}) instead of the Robertson--Messiah relation (\ref{R1}) does not remove problems with the interpretation or derivation of the relation (\ref{M5}) and the characteristic time $\tau_{A} = \infty$ for eigenvectors of $A$ and $H$. It is because for these eigenvectors  relations (\ref{Sch-2}), (\ref{Sch-3}) take also the form of the inequality (\ref{dAdB=0}).

A detailed analysis of  the relation (\ref{H2}) suggests that it may
be in conflict with one of the basic postulates of Quantum Mechanics: Namely, with the projection (reduction) postulate.
It is because the projection postulate leads to the Quantum Zeno Effect \cite{misra} (QZE),  (see also, e.g., \cite{bh,ku,bh1,kk,pf}), that is it makes possible to
force the system to stay in a given state as a result
of continuous or quasi--continuous observations  verifying if the system is in this given state.
It is possible if
the successive measurements (observations) are separated by suitable short time intervals $\Delta t$ such that $\Delta t \to 0$ when the number of observations increases \cite{bh,ku,kk,pf}.
In general  the duration of each of these  measurements
must be shorter than the time interval separating them, and in turn,
the uncertainty of the time $t$ cannot be larger then the duration of these measurements.
Therefore the conclusion  that
the relation (\ref{H2}) should make impossible to observe the QZE seems to be legitimate.
Contrary to such a conclusion there are experimental tests verifying and confirming this effect \cite{WMI}.
The state of the system is characterized by a set of quantum numbers and one of these numbers is the energy of the system in the state considered. Therefore if the quantum system  is forced to stay in the given state by continuously or quasi--continuously checking it if it is in this state, then quantum numbers characterizing this state (including the energy) also remain unchanged. This  means that there is $\Delta E =0$ and $\Delta t \to 0$ in such a case and thus there is a conflict with the relation (\ref{H2}).

The above conclusion can be made to be more reliable by analyzing the conditions guaranteeing the occurrence of the  QZE.
So, let us assume that $|\psi\rangle$ is the state of the system at the initial instant of time $t_{0}$ and let us  analyze the probability ${\cal P}_{\psi}(t_{n}, \ldots ,t_{1},t_{0})$ of finding the system in a given initial state $|\psi\rangle$ in any of measurements performed at instants $ t_{1} < t_{2}< \ldots t_{n}$, ($t_{1} > t_{0}$), which was derived using the projection postulate (see e. q. \cite{ku,bh1}),
\begin{equation}
{\cal P}_{\psi}(t_{n}, \ldots ,t_{1},t_{0}) = \prod_{k=1}^{n}\,|a_{\psi}(t_{k} - t_{k-1})|^{2}. \label{P-n}
\end{equation}
where
\begin{equation}
a_{\psi}(t) = \langle \psi|e^{\textstyle{-i \frac{t}{\hbar}H}}|\psi\rangle. \label{a-psi}
\end{equation}
Now, if to assume for simplicity that $\Delta t = t_{k}  - t_{k-1}$ for all $k=1,2, \ldots, n$, (i. e., that all measurements are separated by equal time intervals),  then the Eq. (\ref{P-n}) takes the following form:
\begin{equation}
{\cal P}_{\psi}(t_{n}, \ldots ,t_{1},t_{0})= |a_{\psi}(\Delta t)|^{2n}.
\label{P-n-Delta}
\end{equation}
The QZE begins to occur when $\Delta t$ is sufficiently small, when $\Delta t \to 0$. So, we need the form of $|a_{\psi}(\Delta t)|^{2}$ for $\Delta t \to 0$.
The analysis of Eq. (\ref{a-psi}) shows that (see \cite{ku})
\begin{equation}
|a_{\psi}(\Delta t)|^{2} \simeq 1 - \left(\frac{\Delta t}{\hbar}\right)^{2} ( \Delta_{\psi}H)^{2} + \ldots,\;\;{\rm for}\;\; t \to 0. \label{a-psi-0}
\end{equation}
This approximate expression correctly describes the short time properties of the square of the modulus of the amplitude $a_{\psi}(\Delta t)$ only if
\begin{equation}
\left(\frac{\Delta t}{\hbar}\right)^{2} ( \Delta_{\psi}H)^{2} \;\ll \;1. \label{zeno}
\end{equation}
Inserting the result (\ref{a-psi-0}) into Eq. (\ref{P-n-Delta}) and taking $\Delta t = \frac{t}{n}$, where $t \equiv t_{n}$, one easily finds that
\begin{equation}
{\cal P}_{\psi}(t, t_{n-1}, \ldots ,t_{1},t_{0})= \left| a_{\psi}(t/n) \right|^{2n} = \left[1 - \left(\frac{t}{n \hbar}\right)^{2} ( \Delta_{\psi}H)^{2}\right]\;\;
\underset{n \to \, \infty}{\longrightarrow} \;\; 1, \label{zeno1}
\end{equation}
which is the Quantum Zeno Effect. Note that the result (\ref{zeno1}) takes place only if the condition (\ref{zeno}) holds. The experimental confirmation of the QZE \cite{WMI} is the proof that the analysis leading to the result (\ref{zeno1}) is correct. What's more, it proves that the assumptions (including (\ref{zeno})) guaranteeing the occurrence of this effect are correct and thus in order to observe the QZE the condition (\ref{zeno}) must be fulfilled. As one can see, the condition (\ref{zeno}) is in direct contradiction to the Heisenberg time--energy uncertainty relation (\ref{H2}): The condition $1 > (\Delta t/\hbar)^{2}\,(\Delta_{\psi}H)^{2} \geq\frac{1}{4}$ is not sufficient for QZE to occur. Therefore the earlier formulated conclusion in this Section that the experimental conformation of the QZE can be considered as the proof that the uncertainty relation (\ref{H2}) can be
in conflict with the projection postulate seems to be justified.

As it was mentioned earlier there is a reservation in \cite{M-T} that derivation of (\ref{M5}) does not go for eigenvectors of $H$ (Then $\Delta H = 0$). In fact it can be only applied for eigenvectors corresponding to the continuous part of the spectrum of $H$. As an example of possible applications of the relation (\ref{M5}) unstable states modeled by wave--packets  of such eigenvectors of $H$ are considered in \cite{M-T}, where using (\ref{M5}) the  relation connecting half--time $\tau_{1/2}$ of the unstable state, say $|\varphi\rangle$, with the uncertainty $\Delta_{\varphi} H$ was found: $\tau_{1/2}\,\cdot\, \Delta_{\varphi} H \geq \frac{\pi}{4}\,h$. In general, when one considers unstable states such a  relation and the similar one appear naturally \cite{fock,kb,boy,grab} but this is quite another situation then that described by the relations (\ref{H1}), (\ref{R1}). The other example is a relation between a life--time $\tau_{\varphi}$ of the system in the unstable state, $|\varphi \rangle$,
and the decay width $\Gamma_{\varphi}$: In such  cases we have $\tau_{\varphi} \cdot \Gamma_{\varphi} = \hbar$ but there are not any uncertainties  of the type  $\Delta E$ and $\Delta t$ in this relation (see, e.g., \cite{fock}). Note that in all such  cases the vector $|\varphi\rangle$ representing the unstable state cannot be the eigenvector of the Hamiltonian $H$. It should be noted here that even in the case
 of unstable states one should be very careful using the relation  (\ref{M5}): For example in the case of unstable states $|\phi\rangle$ modeled by the Breit--Wigner energy density distribution $\omega_{BW}(E) =  \frac{N}{2\pi}\,  \it\Theta (E- E_{min}) \  \frac{{\it\Gamma}_{0}}{(E -E_{0})^{2} +
(\frac{{\it\Gamma}_{0}}{2})^{2}}$,  where $\it\Theta(E)$ is the unit step function and $N$ is the normalization constant, the average values $\langle H\rangle_{\phi} = \int_{E_{min}}^{\infty}\,E\,\omega_{BW}(E)\,dE$ and $\langle H^{2}\rangle_{\phi} = \int_{E_{min}}^{\infty}\,E^{2}\,\omega_{BW}(E)\,dE$ have not definite values and hence $\Delta_{\phi}H$ is undefined which means that the relation (\ref{M5}) does not work in this case.

\section{Conclusions}

In recent years
there has been a growing interest in the uncertainty principles, and in particular the uncertainty principle of time--energy, due to their importance in quantum optics and quantum thermodynamics. So the results presented in this paper seems to be important  not only for the studies of foundations of the quantum theory but also for looking for solutions of some problems in quantum optics and quantum thermodynamics.
Results presented in Sec. 2 allow to draw a conclusion that there can exists such pairs of non--commuting observables  $A$ and $B$ and such vectors that
the lower bound of the product of standard deviations $\Delta A$ and  $\Delta B$ calculated for these vectors is zero: $\Delta A\,\cdot\,\Delta B \geq 0$.
The other conclusion resulting from the analysis prezented in Sec. 2 is that there can also exist such pairs of non--commuting observables $A, B$ and and such complete sets of vectors that the only bound for $\Delta A$ and $\Delta B$, (with $\leq \Delta A < \infty$, $0 \leq \Delta B < \infty$),
calculated for these vectors is $0$.
This means that in such cases restrictions resulting from the uncertainty principle (\ref{R1})  can be bypassed.
So, the Schr\"{o}dinger uncertainty relation (\ref{Sch-2}), (\ref{Sch-3}) and  the Robertson--Messiah relation (\ref{R1}) derived for non--commuting pairs of observables  $A$ and $B$
are  not as universally valid as  it is usually thought.
The discussion of relations (\ref{H2}) and (\ref{M5}) presented in previous Sections
and the detailed analysis of the derivation of the relation (\ref{M5}) suggests that these time--energy uncertainty relations are not well founded and using them one
cannot  consider them as universally valid. Therefore when  using these relations as the basis for predictions of the properties and of a behavior of some systems in physics or astrophysics (including cosmology --- see, e.g., \cite{skr,cos}) one should be very careful interpreting and applying results obtained.
In general in some problems the use of the relation (\ref{M5}) may be reasonable (see, e.g. the case of unstable states)
but  then $\tau_{A}$ used in (\ref{M5}) should not be considered analogously to standard deviations appearing in inequalities  (\ref{H1}), (\ref{R1}).

\section*{Acknowledgments}

The author would like to thank Neelima Kelkar,  Marek Nowakowski and anonymous Referees for  their valuable
comments and discussions.
This work was supported by
the program of the Polish Ministry
of Science and Higher Education under the name "Regional
Initiative of Excellence" in 2019 --- 2022, Project No. 003/RID/2018/19.\\
\hfill\\
{\bf The author contribution statement:} The author declares that there are no conflicts of interest
regarding the publication of this article  and that all results presented in this article are the author's own results.


\end{document}